\documentstyle[12pt]{article}

\def\bq{\begin{quotation}}
\def\eq{\end{quotation}}

\def\fnote#1#2 {\begingroup \def \thefootnote {#1}
\footnote{#2}\addtocounter{footnote}{-1}\endgroup}

\newcommand{\sppt} {\normalsize  Research supported in part by
the Robert A. Welch Foundation and NSF Grant PHY 9009850}

\newcommand{\pagenumber}{\pagestyle{plain}\setcounter{page}{1}}

 \def\G{\Gamma}
\def\a{\alpha} 
\def\b{\beta} 
\def\d{\delta} 
\def\e{\epsilon}

\def\m{\mu}

\def\s{\sigma}

\def\raisenot{\raise .5mm\hbox{/}}
\newcommand{\notpa}{\hbox{{$\partial$}\kern-.54em\hbox{\raisenot}}}
\def\notp{\ \hbox{{$p$}\kern-.43em\hbox{/}}}
\def\notq{\ \hbox{{$q$}\kern-.47em\hbox{/}}}
\def\notk{\ \hbox{{$k$}\kern-.47em\hbox{/}}}
\def\notA{\ \hbox{{$A$}\kern-.47em\hbox{/}}}
\def\nota{\ \hbox{{$a$}\kern-.47em\hbox{/}}}
\def\notb{\ \hbox{{$b$}\kern-.47em\hbox{/}}}

\begin{document}
\baselineskip=24pt

\pagestyle{empty}

\begin{flushright}
UTTG-23-91
\end{flushright}
\begin{center}
{\large \bf
Is $S = 1$ for $c=1$?}\fnote{1}{\sppt} \\
\vspace{0.3in}
Djordje Minic\\
{Zhu Yang}\fnote{2}{\normalsize
Adress after September 1, 1991: Department of Physics
and Astronomy, University of Rochester, Rochester, NY 14627}\\

\vspace{0.2in}
Theory Group\\
Department of Physics\\
University of Texas\\
Austin, TX 78712\\
\vspace{0.35in}
{\large \bf
ABSTRACT}
\vspace{0.2in}
\end{center}

The $c=1$ string in the Liouville field theory approach is shown to possess
a nontrivial tree-level $S$-matrix which satisfies factorization property
implied by unitary, if all the extra massive physical states are included.

\newpage
\pagenumber

String amplitudes and their factorization properties in the Liouville
field theory \cite{sb,joe} are not yet
well understood at present. In this note we
show that the spherical
$c=1$ string amplitudes computed from the Liouville theory according to
conventional vertex operator normalization
are nontrivial and satisfy correct factorization.
The renormalized amplitudes computed by
Gross and Klebanov \cite{gros} imply
that $S=1$. This result agrees with the so-called ``bulk" $S$-matrix
in the collective field theory of $c=1$ quantum gravity.
We, on the other hand,  follow Polyakov \cite{pol} who considered
unrenormalized amplitudes and find a unitary $S$-matrix
not equal to 1.
The $S$-matrix we are considering is also different from the ones computed in
matrix models \cite{mat} (see also [6,7]),
which are ``wall"  amplitudes.
As a necessary byproduct of our analysis, a host of extra states appears in
the spectrum of the theory. These states are not present in the naive
light-cone analysis of the spectrum (according to this
there is only one degree of freedom, the center of mass of the string,
so-called ``tachyon"), but appear naturally both in factorization of amplitudes
as intermediate states and in the pole structure of the corresponding
world-sheet operator product expansion ($OPE$).

We hope that our results will shed some light on difficult issues such as
factorization and nature of spectrum in the Liouville theory.
Due to the apparent difference of our results and conclusions on [3], a
subtle relation is to be expected between the continuous collective field
theory \`{a} la Das-Jevicki and the related Liouville theory of $c=1$ quantum
gravity.

The organizaton of this note is as follows. First, we review some well-known
facts about $c=1$ string theory. Then we discuss factorization of amplitudes
at tree level. Lastly, we concentrate on the $OPE$'s of the underlying
world-sheet theory in attempt to understand the pole structure of the
factorization formulae.

In the path integral approach to string theory, the S-matrix
is computed from sum over surfaces with insertion of local vertex
operators of correct dimensions. One checks for unitarity to verify the
consistency of the result. At tree level, the unitarity reduces to
factorization of amplitudes \cite{wein}. The underlying reason is of course
the $OPE$ of the world-sheet conformal field theory.
A singularity in the OPE \cite{joe2} is directly related to the pole structure
in the factorization
formula. Indeed, in their original computation of the 2 point function in
c=1 string theory, Gross, Klebanov and Newman \cite{gros2}
have derived some OPEs and observed
their pole structure.
In light of their analysis we consider later on the $OPE$s
of the world-sheet theory and find signs of a nontrivial spectrum and
$S$-matrix.

Let us start from the general expression for the $N$-point tachyon
amplitude of
 the $c=1$  string with  the zero cosmological constant
in the path integral approach \cite{pol2,ddk}:
\begin{eqnarray}
A_{N} &=& \int DtD\phi\, \prod^{N}_{i=1}\int d^{2}z_{i}\,\sqrt{\hat{g}(z_{i})}
V_{\b_{i},p_{i}}(z_{i})\, \nonumber \\
&~& \exp\,\lbrace-{\frac{1}{8\pi}}
\int d^{2}\s \sqrt{\hat{g}}\,[\,\partial_{\a}t\,\partial^{\a}t+
\partial_{\a}\phi\,
\partial^{\a}\phi-2\sqrt{2}\,\hat{R}\,\phi\,]\rbrace.
\end{eqnarray}
Here the vertex operator $V_{\b_{i},p_{i}}$ creates
a massless ``tachyon" at momentum
$p_{i}$ with chirality $\e(i)=sgn (p_{i})$,
$t$ represents Euclidean time, and $\phi$ is
understood as a spatial coordinate. Conformal invariance dictates the form
of the vertex operator
\begin{eqnarray}
V_{\b_{i},p_{i}}(z_{i}) &=& \exp\, (ip_{i}t(z_{i})-\sqrt{2}\phi(z_{i})+|p_{i}|
\phi(z_{i}))\, \nonumber \\
&\equiv& \exp\, (ip_{i}t(z_{i})+\b_{i}\phi(z_{i})).
\end{eqnarray}
Apart from some non-covariant factors, (1) defines transition amplitudes
in string theory.
Evaluation of the corresponding path integral in $D=26$ results in the
well-known Virasoro-Shapiro amplitude \cite{vis}.

In what follows kinematics will play a major role. There exists a standard
procedure that leads to the peculiar kinematical constraints in the Liouville
theory with the Euclidean signature. One integrates over the zero modes of $t$
and $\phi$ in (1) ($t_{0}$ and $\phi_{0}$ respectively).
Integration over $t_{0}$ gives the momentum conservation law. Integration
over $\phi_{0}$ is not well-defined, so one rotates $\phi_{0}$ to $i\phi_{0}$
and deduces an ``energy" conservation law. All in all one is left with [3,4],
\begin{equation}
\sum^{N}_{i=1} p_{i}=0,
\hspace{0.1in} \sum_{i=1}^{N} |p_{i}| =(N-2) \sqrt{2}.
\end{equation}
We prefer to think about physical scattering in 2D Minkowski
space-time\footnote{The following picture
was suggested to us by J. Polchinski.}.
Consider, for example, a three-point tachyon amplitude.
For that purpose, prepare two suitably normalized wave packets of tachyons
and let them collide. This is of course a well-defined physical process:
insertion of
wave packets should render previously divergent integrals analytically
well behaved. Indeed, if the initial state is represented as
\begin{equation}
|i\rangle \sim \int dp_{1} dp_{2} f_{1}(p_{1}) f_{2}(p_{2})
|p_{1},p_{2},in\rangle,
\end{equation}
where $p_{1}$ and $p_{2}$ are incoming momenta and $f_{1}(p_{1}), f_{2}(p_{2})$
are peaked around the actual momenta of the incident particles with a
finite and small width, and the final state is approximated by a plane
wave, integration over $t_{0}$ implies momentum conservation.
On the other hand, integration over $\phi_{0}$ is now well-defined with
the corresponding integral
\begin{equation}
\sim \int d\phi_{0} dp_{1} dp_{2} f_{1}(p_{1}) f_{2}(p_{2})
\exp [i(p_{1}+p_{2}-p_{3}
+i\sqrt{2})\phi_{0}],
\end{equation}
leading to the ``energy sum rule" $p_{1}+p_{2}-p_{3}=-i\sqrt{2}$
in Minkowski space-time. Therefore, the formal precedure based on the
Wick rotation of $\phi_{0}$
is supported by a reasonable physical picture, and leads to the same result.

In what follows we will use $A_{m,N-m}$ to denote an $N$-point amplitude
(1) of $m$ points with $+$ chirality and $(N-m)$ points with $-$  chirality.
The evaluation of (1) has been done in \cite{kuta,gros,pol}, where
the integrals calculated in \cite{dos} were used.
The integral representation of (1) is
\begin{eqnarray}
A(1,2,\cdots,N) &=& \int \prod^{N}_{i=1}d^{2}z_{i} \,\m(z_{i})\,
\langle \prod^{N}_{i=1}
V_{\b_{i},p_{i}}\rangle\, \nonumber \\
&=& \int \prod^{N-3}_{i=1}d^{2}z_{i} |z_{i}|^{-2s_{i\,N-2}}|1-z_{i}|^{-2s_{i\,
N-1}} \prod_{i<j} |z_{i}-z_{j}|
^{-2s_{ij}},
\end{eqnarray}
where $s_{ij}=(-\sqrt{2}+|p_{i}|)(-\sqrt{2}+|p_{j}|)-p_{i}p_{j}$ and
``$\langle \rangle$" means free field contraction. The function $\m(z_{i})$
is an appropriate Faddeev-Popov determinant from the $SL(2,C)$ gauge fixing.
In particular \cite{dos,gros,dif},
\begin{eqnarray}
A_{2,1}&=& A_{1,2}=1,\\ \nonumber
A_{3,1}&=&
\pi\prod_{i=1}^{3}{\frac{\G(1-\sqrt{2}p_{i})}{\G(\sqrt{2}p_{i})}},\,\
,
\\ \nonumber
A_{N-1,1} &=& {\frac{\pi^{N-3}}{(N-3)!}}
\prod_{i=1}^{N-1}{\frac{\G(1-\sqrt{2}p_{i})}
{\G(\sqrt{2}p_{i})}}.\,
\end{eqnarray}
The last formula can be established if one uses symmetries of the integrand
in (6) in conjunction with convenient kinematical constraints and makes
an ansatz such that successive reduction from the $(N-1,1)$ case leads to
the 4-point function, for which the analytic expression can be explicitly
written.

We illustrate the outlined procedure for $A_{4,1}$ amplitude. In this case,
(6) is invariant under the following substitutions:
\begin{eqnarray}
z_{1} &\rightarrow& 1-u_{1}, \,\, s_{13}\rightarrow s_{14} \, \\ \nonumber
z_{2} &\rightarrow& 1-u_{2}, \,\, s_{23}\rightarrow s_{23}-s_{13}+s_{14},
\end{eqnarray}
where $s_{ij}=2-\sqrt{2}(p_{i}+p_{j})$, $p_{i}\ge 0$ for $i=1,\cdots,4$,
and $p_{5}=-3/\sqrt{2}$. Also $s_{24}=s_{23}-s_{13}+s_{14}$ as a consistency
requirement. Then by writing $A_{4,1}$ as
\begin{eqnarray}
A_{4,1} &=& T({\frac{2-s_{12}-s_{13}+s_{23}}{2}})
T({\frac{2-s_{21}-s_{23}+s_{13}}{2}})\, \\ \nonumber
&~& T({\frac{2-s_{31}-s_{23}+s_{12}}{2}})
T({\frac{2+s_{12}+s_{13}-s_{23}-2s_{14}}{2}})f(s_{13},s_{14},s_{23},s_{12}),
\end{eqnarray}
where $T(x)=\G(1-x)/\G(x)$, one can see that $f(s_{13},s_{14},s_{23},s_{12})$
has to satisfy the above symmetry relations.
Following Dotsenko and Fateev \cite{dos}, one proves that
$f$ is a bounded and analytic function of $s_{ij}$ in the entire complex plane.
Then $f$ is a constant that can be deduced by truncating $A_{4,1}$ to
$A_{3,1}$.
Apparently the same procedure persists for higher order correlation functions
of the form $(N-1,1)$ \cite{dif}.
Observe that from the expression for $A_{N-1,1}$ one can get $A_{N-2,2}$
by picking an appropriate value for one of the (N-1) momenta, and so on
for $A_{M-m,m}$.
That means that
generically  $A_{N-m,m}=0$ if $m=2,\cdots, N-2$. We illustrate this by
displaying $A_{2,2}$,
\begin{equation}
A_{2,2}=\pi{\frac{\G(1-s_{14})}{\G(s_{14})}}{\frac{\G(1-s_{24})}{\G(s_{24})}}
{\frac{\G(1-s_{34})}{\G(s_{34})}},
\end{equation}
where $p_{1,2}\ge 0$ and $p_{3,4}\le 0$. The conservation laws imply (3) that
$p_{1}+p_{2}=-(p_{3}+p_{4})=\sqrt{2}$, thus $s_{34}=0$ and $A_{2,2}=0$.
It is clear that $A_{N-1,1}$ has poles for exceptional values of the momenta
$p_{i}=(M+1)/\sqrt{2}, M=0,1,\cdots$.
If we Wick-rotate back to the Minkowski signature $
(p\rightarrow ip)$, the poles are at
imaginary momenta.
It has
been suggested  \cite{gros} that one should simply absorb them
through a wave function
normalization. After the renormalization all amplitudes vanish, giving
$S=1$ for $c=1$. Such an S-matrix is
of course trivially unitary. Here we propose
an alternative interpretation of (1), which leads to a tree level unitary,
yet non-trivial S-matrix. First we recall what is meant by
tree-level factorization (see, for example, \cite{wein,joe2}
where the correct vertex operator normalization is derived from tree-level
factorization).
When the total momentum $p_{\m}$ of some set of external legs,
say 1,2, $\cdots$, $L$, approaches
the mass shell of a physical particle of mass $m$ and type $j$,
 the $S$-matrix must have a pole with
\begin{equation}
A(1,2,\cdots, N)=-i(p^{2}+m^{2})^{-1}\,\sum_{j}A(1,2,\cdots,L,j)\,
A(j,L+1, L+2, \cdots, N),
\end{equation}
where $p\equiv p_{1}+\cdots+p_{L}$.
There are only minor changes in the above equation when applied to
our situation.
Firstly, since the tachyon is really massless
in $c=1$ string theory we have $m=0$. More importantly, in
string theory, the propagator is
in general defined as $(L_{0}-1)^{-1}$. For an off-shell
tachyon vertex operator $\int
d^{2}\sigma \,\sqrt{\hat{g}}\, \exp\,(ipt-\sqrt{2}\phi\pm q \phi)$, $L_{0}$ is
$1+p^{2}-q^{2}$. So the tachyon propagator is just $(p^{2}-q^{2})^{-1}$.
Finally, as we will see,
a satisfactory factorization
must include the extra states discovered in \cite{pol} in the physical
Hilbert space.

Before investigating the general case, let's work out the factorization of (1)
for the 2 particle scattering amplitude $A_{3,1}$.
We assume that $p_{3}+p_{4}+p=0, p_{3}-p_{4}-q=\sqrt{2}$, i.e., the off-shell
intermediate state in the $s$ channel (figure 1)
carries 2-momentum $(p, -\sqrt{2}-q)$.
For the $(3,1)$ kinematic region, $p_{4}$ is easily worked out as
$p_{4}=-\sqrt{2}$. So we can express $p_{3}$ as $p_{3}=-p+\sqrt{2}$.
The on-shell condition of the intermediate state is then $p=q=1/\sqrt{2}$,
which also implies that $p_{2}=1/\sqrt{2}-p_{1}$.
Now we can rewrite $A_{3,1}$ as
\begin{equation}
A_{3,1}=\pi {\frac{\G(1-\sqrt{2}p_{1})}{\G(\sqrt{2}p_{1})}}
{\frac{\G(1-\sqrt{2}p_{2})}{\G(\sqrt{2}p_{2})}}
{\frac{\G(\sqrt{2}p-1)}{\G(2-\sqrt{2}p)}}.
\end{equation}
As $p\rightarrow 1/\sqrt{2}, \G(\sqrt{2}p-1)\rightarrow (\sqrt{2}p-1)^{-1}$,
and the last $\G$ function
in the denominator approaches 1. (8) then leads to
\begin{eqnarray}
A_{3,1} &\rightarrow& \pi {\frac{\G(1-\sqrt{2}p_{1})}{\G(\sqrt{2}p_{1})}}
{\frac{\G(1-\sqrt{2}(1/\sqrt{2}-p_{1})}{\G(\sqrt{2}(1/\sqrt{2}-p_{1})}}
{\frac{1}{\sqrt{2}p-1}}\\
\nonumber
&\rightarrow& {\frac{\pi}{p^{2}-q^{2}}}.
\end{eqnarray}
Since $A_{2,1}=A_{1,2}=1$, (9) satisfies the correct factorization.

It appears paradoxical that $A_{2,2}=0$, since
the unitarity requires
that $A_{2,2}\ne 0$.
For example, in the situation indicated in figure 2, when the intermediate
state is close to the tachyon mass-shell $p+q\sim 0$,
\begin{equation}
A_{2,2} \rightarrow {\frac{1}{p^{2}-q^{2}}}.
\end{equation}
The resoluton is also simple. If the intermediate state is on the mass-shell,
the kinematical relation tells us that $p_{1}=p_{2}=1/\sqrt{2}$.
One can easily work out that $s_{1,4}=s_{2,4}=1$, so
\begin{equation}
A_{2,2} = \pi {\frac{\G(1)}{\G(0)}}{\frac{\G(0)}{\G(1)}}
{\frac{\G(0)}{\G(1)}}.
\end{equation}
To be sure, (15) really means a Dirac $\d$-function at $p_{1}=1/\sqrt{2}$.
Indeed, after Wick rotation with  suitable $i\e$ prescription, one can show
that $A_{2,2}$ is $i \d (p_{1}-1/\sqrt{2})$.
Thus unitarity holds in a strange way. Clearly, many analytical
properties familiar in the case of a four dimensional $S$-matrix
are lost here.

Generalization to $A(N-1,1)$ is easy (figure 3). The factorization
we are interested in is $A(N-1,1)\rightarrow A(N-2,1)A(2,1)$.
Again we look for an on-shell pole at total momenta
$p_{N-1}+p_{N}+p=0, p_{N-1}-p_{N}-q=\sqrt{2}$. Since $p_{N}$ can be worked out
to be $(2-N)/\sqrt{2}$, the on-shell condition is $p=q=(N-3)/\sqrt{2}$.
Using the kinematic relation we can express $p_{N-1}$ as $(N-2)/\sqrt{2}-p$.
Substituting this into (4) we find, near the mass shell
\begin{eqnarray}
A_{N-1,1} &\rightarrow& {\frac{\pi^{N-3}}{(N-3)!}}
\prod_{i=1}^{N-2}{\frac{\G(1-\sqrt{2}p_{i})}
{\G(\sqrt{2}p_{i})}} {\frac{1}{\sqrt{2}p-(N-3)}}\\ \nonumber
&\rightarrow& {\frac{\pi^{N-3}}{(N-4)!}}
\prod_{i=1}^{N-2}{\frac{\G(1-\sqrt{2}p_{i})}
{\G(\sqrt{2}p_{i})}} {\frac{1}{p^{2}-q^{2}}}\\ \nonumber
&\rightarrow& {\frac{\pi}{p^{2}-q^{2}}}A_{N-2,1}A_{1,2},
\end{eqnarray}
which is again what one would expect from factorizability.
Note that although the pole in (8) corresponds to a resonance of high
momentum $p=(N-3)/\sqrt{2}$ of the intermediate state, it is the lowest
pole in external legs. The resonant state is the ordinary massless tachyon
with peculiar values of its momenta. We will discuss higher order poles
in the external legs later.

The tachyon poles at discrete momenta can also be observed in the structure of
the OPE in the underlying ``gravitationally dressed" conformal field theory.
This should be expected on general grounds \cite{joe2}.
Consider the same $N$ tachyon amplitude as in the last paragraph.
{}From (4), if we fuse the last two vertex operator using  the free field
$OPE$, we obtain for the first short distance singularity ($z=z_{N}-z_{N-1}$),
\begin{eqnarray}
A_{N-1,1} &\sim& \int d^{2}z |z|^{-4+2\sqrt{2}(p_{N-1}-p_{N})
+4p_{N-1}p_{N}}\int
\prod^{N-1}_{i=1} d^{2}z_{i}\,\m(z_{i}) \, \nonumber \\
&~&\langle V_{\b_{1},p_{1}}(z_{1}) \cdots V_{\b_{N-1}+\b_{N},
p_{N-1}+p_{N}}(z_{N-1})\rangle \, \nonumber \\
&\sim& {\frac{1}{\sqrt{2}p-(N-3)}} \int
\prod_{i=1}^{N-1}d^{2}z_{i} \m(z_{i}) \langle V_{\b_{1},p_{1}}(z_{1})
\cdots V_{\b_{N-1}+\b_{N},
p_{N-1}+p_{N}}(z_{N-1})\rangle,
\end{eqnarray}
where we have used kinematic relation $p_{N}=(2-N)/\sqrt{2}$ and $p_{N-1}
+p_{N}+p=0$.
Indeed, the lowest
tachyon pole which appears in the factorization of the $N$-point amplitude in
the Liouville theory (eq.(9)) can be clearly seen.

Now we offer a similar explanation for higher poles in external leg as exchange
of extra states \cite{pol}. Take, for example, a 4-point amplitude  in (6) and
work out the free field $OPE$ of $V_{\b_{3}, p_{3}}$ and $V_{\b_{4},p_{4}}$,
\begin{equation}
V_{\b_{3}p_{3}}(z_{3},\bar{z_{3}})V_{\b_{4}p_{4}}(z_{3}+z,\bar{z_{3}}+\bar{z})
=\sum_{M,\bar{M}}|z|^{-4+2\sqrt{2}(p_{3}-p_{4})+4p_{3}p_{4}}z^{M}\bar{z}^{
\bar{M}} O_{M}^{4}(z_{3})O_{\bar{M}}^{4}(\bar{z_{3}}),
\end{equation}
where
\begin{equation}
O^{4}_{M}(z_{3})={\frac{1}{M!}}:\exp [\b_{3}\phi(z_{3})+ip_{3}t(z_{3})]
\partial^{M}_{z} \exp [\b_{4}\phi(z_{3})+ip_{4}t(z_{3})]:
\end{equation}
and  similarly for the antiholomorphic operator $O^{4}_{\bar{M}}(\bar{z_{3}})$.
By inserting the $OPE$ into (6) and using the kinematical relations $\b_{4}=0$
and $p_{4}=-\sqrt{2}$, we obtain the singular contribution
\begin{eqnarray}
A_{3,1} &\sim& \sum_{M,\bar{M}}
 \int d^{2}z |z|^{-4+2\sqrt{2}(p_{3}-p_{4})
+4p_{3}p_{4}}z^{M}\bar{z}^{\bar{M}}
 \int \prod_{i=1}^{3} d^{2}z_{i} \,\m(z_{i})\, \nonumber \\
&~& \langle V_{\b_{1},p_{1}}(z_{1}) V_{\b_{2},p_{2}}(z_{2})\,
 O^{4}_{M}(z_{3}) O^{4}_{\bar{M}}(\bar{z_{3}})  \rangle\,
\nonumber \\
&\rightarrow& \sum_{M}
{\frac{1}{\sqrt{2}p_{3}-(M+1)}} \int \prod^{3}_{i=1} d^{2}z_{i}
\m(z_{i})
\langle V_{\b_{1},p_{1}} V_{\b_{2},p_{2}} \,
O^{4}_{M}(z_{3}) O^{4}_{\bar{M}}(\bar{z_{3}})\rangle.
\end{eqnarray}
Clearly the external leg poles
in (7) correspond to $A_{3,1}$ factorizing into
$O^{4}_{M}(z_{3}) O^{4}_{\bar{M}}(\bar{z_{3}})$ at special momenta
$p_{3}=(M+1)/\sqrt{2}$, and $O^{4}_{M}(z)$ takes the form
\begin{equation}
O^{4}_{M}(z_{3})={\frac{1}{M!}}:\exp [(M-1)\phi(z_{3})/\sqrt{2}+
i(M+1)t(z_{3})/\sqrt{2}]
\partial^{M}_{z} \exp [-i\sqrt{2}t(z_{3})]:.
\end{equation}
These are precisely the special operators considered by Danielsson and
Gross \cite{dan} in the $c=1$ conformal field theory.
Generalization to $A_{N-1,1}$ presents no difficulty. We will find
that even more special operators appear. Now let us fuse
$V_{\b_{N-1}p_{N-1}}$ and $V_{\b_{N}p_{N}}$ in
$A_{N-1,1}$ and integrate over $z$ and $\bar{z}$. We obtain
\begin{equation}
A_{N-1,1} \sim  \sum_{M}
{\frac{1}{\sqrt{2}p_{N-1}-(M+1)}} \int \prod^{N-1}_{i=1} d^{2}z_{i}
\m(z_{i})
\langle V_{\b_{1},p_{1}} \cdots V_{\b_{N-1},p_{N-1}} \,
O^{N}_{M}(z_{N-1}) O^{M}_{\bar{M}}(\bar{z_{N-1}})\rangle,
\end{equation}
where
\begin{equation}
O^{N}_{M}(z_{N-1})={\frac{1}{[M(N-3)]!}}: \exp[\b_{N-1}\phi+
ip_{N-1}t]
\partial^{M(N-3)}_{z} \exp [\b_{N}\phi+ip_{N}t]:,
\end{equation}
and $\b_{N}=(N-4)/\sqrt{2}, p_{N}=(2-N)/\sqrt{2}$. In deriving (23)
we have taken into account the fact that some poles in the $OPE$ do not
show up in the final answer (7). When $p_{N-1}=(M+1)/\sqrt{2}$,
we get the following special operators
\begin{equation}
O^{N}_{M}={\frac{1}{[M(N-3)]!}}:\exp [(M-1)\phi/\sqrt{2}+
i(M+1)t/\sqrt{2}]
\partial^{M(N-3)}_{z} \exp [(N-4)\phi/\sqrt{2}+i(2-N)t/\sqrt{2}]:.
\end{equation}
Unlike $O^{4}_{M}$, $O^{N}_{M}$ ($N>4$) mixes $\phi$ and $t$ in a nontrivial
way. However, simple arguments \cite{hwang} show that $O^{N}_{M}$ can be
reduced to a product of a nontrivial $t$ primary field with a pure
exponential operator of $\phi$, plus spurious operators.
By the way, looking at the propagator corresponding to the special
operators, one finds that they describe massive intermediate states.

Since extra massive states exist in the intermediate channel, they should
also appear as external states for the reason of unitarity.
One then has to consider factorization
of amplitudes involving them. These states appear only at discrete momenta,
thus the physical picture of their scattering is not very clear.

Recently a number of authors \cite{dan,gm,sen} have considered extra
degrees of freedom in $c=1$ string theory from different points of
view. Comparing our results with \cite{dan} we find an agreement.
Their analysis was done from the point of view of matrix models which
indicates a more precise relation between the Liouville theory and
matrix models. Closely related is the work of \cite{gm} where the
issue of extra states was discussed directly in the Liouville theory.
We would like to point out that our $S$-matrix obviously does not agree with
the so-called ``bulk" $S$-matrix calculated in the
collective string field theory
$\cite{das}$ by Gross and Klebanov $\cite{gros}$. Therefore it is not
clear how
the collective string field theory exhibits stringy degrees
of freedom found in the Liouville theory.

In conclusion, we have shown that the naive definition of string amplitudes
gives rise to a non-trivial $S$-matrix with the desired property of
factorization, if we take into account the extra physical states.
The nature of extra states is obscure though. It is not obvious to us
that they can be related to topological degrees of freedom of $D<1$
string theory \cite{pol}.

\vspace{0.2in}
\begin{flushleft} {\large \bf Acknowledgement}
\end{flushleft}
We are grateful to Joe Polchinski for suggesting to us to look at factorization
of amplitudes in $c=1$ string theory. His inspiration, insight and constant
interest were of inestimable importance to us.

\end{document}